# A Novel C2C E-Commerce Recommender System Based on Link Prediction: Applying Social Network Analysis

Mohammad Dehghan Bahabadi
Department of Computer Engineering and Information Technology
Amirkabir University of Technology
Tehran, Iran
Dehghan.Bahabadi@aut.ac.ir

Alireza Hashemi Golpayegani
Department of Computer Engineering and Information Technology
Amirkabir University of Technology
Tehran, Iran
Sa.Hashemi@aut.ac.ir

Leila Esmaeili
Department of Computer Engineering and Information Technology
Amirkabir University of Technology
Tehran, Iran
Leila.Esmaeili@aut.ac.ir

*Abstract*— **Social network analysis emerged as an important research topic in sociology decades ago, and it has also attracted scientists from various fields of study like psychology, anthropology, geography and economics. In recent years, a significant number of researches has been conducted on using social network analysis to design e-commerce recommender systems. Most of the current recommender systems are designed for B2C e-commerce websites. This paper focuses on building a recommendation algorithm for C2C e-commerce business model by considering special features of C2C e-commerce websites. In this paper, we consider users and their transactions as a network; by this mapping, link prediction technique which is an important task in social network analysis could be used to build the recommender system. The proposed tow-level recommendation algorithm, rather than topology of the network, uses nodes' features like: category of items, ratings of users, and reputation of sellers. The results show that the proposed model can be used to predict a portion of future trades between users in a C2C commercial network.**

*Keywords*— *Recommender System; Commercial Network; Link Prediction; C2C Commerce; Social Network Analysis*

## I. INTRODUCTION

In recent years, moving from mass production toward mass customization has been an important factor for success of companies in global markets. In other words, diversity and customization of products have replaced long life time and standardized products. The emersion of electronic commerce has facilitated their movement toward personalization of products. In order to expand personalization, companies need to increase the amount of information that users should process before choosing their ideal items. This phenomenon which is known as Information overload problem, was the main reason of developing recommender systems in e-commerce [1].

Recommendation based on best-selling items, demographic information of users, and user's past behavior analysis, are samples of techniques used by Recommender systems in e-commerce websites [2]. These techniques can be considered as different types of personalization, since they help the website to adapt itself to the customers. Although most of commercial websites use B2C e-commerce, some of them support C2C e-commerce. Ebay.com, Bizrate.com, etc. are instances of websites in which customers can be buyer and seller at the same time. Current recommendation algorithms (e.g. collaborative filtering, content based algorithms, and hybrid algorithms) can be most useful for B2C e-commerce websites. Since these algorithms only concentrate on items and consumers (but not sellers), there is a lack of recommender systems for C2C e-commerce websites [3].

The solution proposed in this paper for the mentioned problem is based on social network analysis. Social network analysis emerged as an important research topic in sociology decades ago, and its first studies was focused on the adoption of medical and agricultural innovations [4]. It is a vast field of research that has attracted researchers from anthropology, economics, psychology, biology and geography, just to mention a few. Link prediction which is an important task in social network analysis, is the problem of predicting the existence of a link between two nodes in a graph where prediction is based on the attributes of the nodes and other observed links [5].

In this paper, we are going to design a recommender system which can be used in commercial networks. A commercial network is a graph which its nodes represent users of a C2C e-commerce website and edges represent the transactions between these users. If we presume transactions between users as edges of a graph, link prediction approaches can be used to design the recommender system. But before using link prediction techniques, the nature of network should be identified and analyzed. Since the context of links in commercial networks are different from links in common social networks, we must adapt link prediction approaches to our problem. So far, several nodes proximity measures are introduced for link prediction in common social networks (e.g. Adamic-Adar, FOAF, etc. [6]) but as we will discuss in section III, many of them are not applicable for being used in our model.

The main goal of this paper is to propose a model for predicting future trades between users in a commercial network. If $G_I(U,E_I)$ is the transactions graph in time $t_I$ ($U$ is the set of nodes and $E_I$ is the set of graph edges in time $t_I$). We are going to predict edges of the graph in time $t_2$ ($t_I<t_2$). So we should predict the graph $G_2(U,E_2)$ in time $t_2$ where the difference between $G_I$ and $G_2$ is in their edges. Predicting these new links and using them in a recommender system could have two benefits: first, such recommender system can accelerate formation of trades between users and this can benefit all involving parties, i.e. buyers and sellers and third parties. Second, recommending a user to buy an item from another user can end up to a trade that was not going to happen in future at all. Another goal that we are going to seek is to predict items sold in future trades. Since the ultimate goal of this paper is recommending items of sellers to target users, we should choose the most appropriate item of the sellers to increase the probability of accepting recommendations.

The rest of the paper is organized as follows: In section II, we briefly review some of previous works on link prediction problem in social networks and recommendation algorithms for C2C websites. Section III contains our proposed methodology for solving the problem ahead. In section IV we discuss the experiments we performed and the data used in experiments. Results are presented in section V, and finally in section VI we present conclusions and future works.

## II. RELATED WORK

So far, plenty of research has been done in the field of link prediction and item recommendation in common social networks, but just a few of them has focused on commercial networks. In [7], a recommendation model for C2C online trading is proposed. The author divides user's behaviors in a C2C website into four categories: browse, attention, bidding, and purchase. Based on these criteria, a users' preference matrix is built. Afterwards, similarity between current active users and other users is calculated. Finally Top-N recommendation list is built for each user based on most important items for his neighbors. Three-dimensional collaborative filtering is another method for recommending items in C2C commercial networks which is introduced in [8]; They extended the traditional collaborative filtering technique to solve the recommendation problem in C2C e-commerce websites. Their proposed recommender system firstly calculates seller similarities using seller features, and fills the rating matrix based on seller similarities. Then it calculates the buyer similarities using historical ratings, and defines neighbors of each user to predict unknown ratings. Eventually, the system recommends the seller and product combinations with the highest prediction ratings to the target user.

In, Arazy et al. [9] proposed a framework for improving social recommender systems using behavioral theories of advice-taking. They identified four factors that impact a user's decision when he receives a recommendation: Homophily, Trust, Reputation, and Tie strength. After computing these relationship factors, the system should calculate a weighted average of them in order to rate each recommendation source. Then these sources could be used to recommend items to users. In [10], Guy et al. studied personalized recommendation of social software items, including bookmarked web-pages, blog entries, and communities. They focused on recommendations that are derived from the user's social network. In their proposed model, Social network information is collected and aggregated across different data sources. A group recommendation system is proposed in [11]. The system makes recommendations based on supervised entropy, association rules and D-Tree classification method.

Unlike item recommendation in C2C commercial networks, a large number of researches has been done on link prediction and link recommendation in common social networks. Some works have used only network's topology for predicting new links. Nowell et al. in [5], state that link prediction could be done just by using topological features. They used three different methods to predict missing links: Methods Based on Node Neighborhoods, Methods Based on the Ensemble of All Paths, and Higher Level Approaches. In [6], de Sá et al. investigated the effect of using supervised learning approaches on weighted graphs. They used pairs of nodes as instances of binary classification problem in which the predictor attributes were metrics computed from the network topology, and the target attribute was the presence or absence of a link between the nodes in future. Link prediction using hybrid approaches is investigated in many of prior works. Yin et al. [12] proposed a framework for link recommendation based on attribute and structural properties in a social network. They enumerated six criteria for link recommendation: homophily, rarity, Social influence, common friendship, social closeness, and preferential attachment. In [13], Chen et al. compared the performance of different people recommendation in predicting and recommending potential friends. The comparison was between four algorithms: content matching, content plus link, friend of a friend, and SONAR.

To the best of our knowledge, except [8], none of the previous researches about item recommendation in C2C e-commerce websites are focused on choosing appropriate seller for target users, and this is one of the main difference between this work and previous works. Besides, this paper uses link prediction technique for recommending items in C2C e-commerce websites, an approach that has not been used in previous researches about C2C e-commerce recommender system.

## III. PROPOSED MODEL

In general, link prediction approaches are divided into three categories: methods based on structural measures in the network, methods based on the content or attribute similarity between nodes, and methods based on probabilistic models. In this paper, measures based on structure of network and attributes of nodes are used for link prediction task. This choice is according to the results of previous works which state the more features of social network we use, the stronger predictions we can make [13], [12]. Our proposed model for link recommendation (i.e. recommendation of buying items from sellers) in commercial networks is shown in Fig. 1. These are the assumptions that we made to design the model:

- There is a network of users in which some users are just seller, some are just buyer, and some are both.

- After every transaction, buyers can rate sellers according to four aspects: overall satisfaction, quality of goods, delivery time, and seller support.

- Each item belongs to a certain category (number of categories are limited).

As you can see in Fig. 1, the recommendation model consists of six stages. It should be noted that stage 2, 3 and 4 can be done in parallel, but other stages must be performed in specified order.

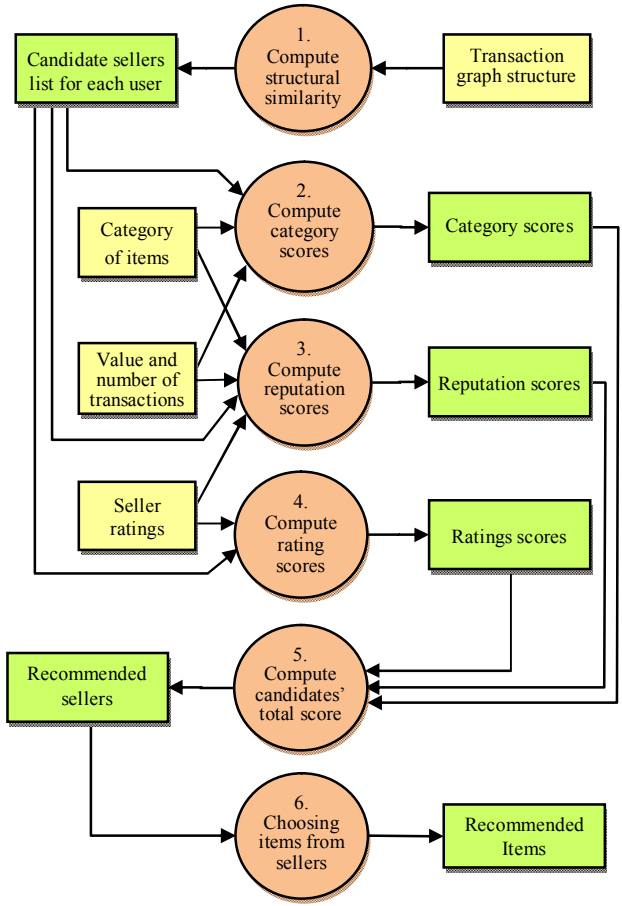

Fig. 1. *Proposed Link Recommendation Model for C2C Commercial Networks. Rectangles represent input (yellow) or output (green) data, and circles represent system processes.*

## A. Computing structural similarity

The first stage of our model is computing similarity of each pair of nodes according to network structural metrics. Considering definition of the problem, we cannot directly use these metrics. Because unlike common social networks, there is no clear evidence that having more common neighbors or being close to each other in graph, increases the probability of forming a link between two nodes. So in this stage, a different approach must be used. SimRank is a similarity measures proposed for directed graphs. This measure which is proposed by Jeh & Widom in [14], computes similarities based on an intuitive concept: two objects are similar if they refer to similar objects or they are referred by similar objects. Like PageRank algorithm [15], SimRank computes similarity of graph nodes using their input and output edges. The first step is to calculate user similarities according to SimRank measure. We use Eq. (1) to calculate similarity of target users to other users in transactions graph.

$$S(u,v) = \frac{C}{|I(u)\|I(v)|} \sum_{i=1}^{|I(u)|} \sum_{j=1}^{|I(v)|} S(I_i(u), I_j(v)) \quad (1)$$

In Eq. (1), $u$ and $v$ are sample users, $S(u,v)$ is the similarity of $u$ and $v$, $C$ is the damping factor, $I(u)$ is the list of sellers that user $u$ have purchased items from and $I(v)$ is the list of sellers that user $v$ have purchased items from. After computing similarity of each target user like $u$ to other users in graph, there will be a list of $n$ similar users to $u$. We use the list of similar users to specify candidate sellers; for this purpose, we choose sellers that have sold items to users in $u$'s similarity list (but $u$ has bought nothing from them) as candidate sellers. Assume the graph in Fig. 2 as a commercial network. Nodes on the left are buyers and nodes on left are sellers. Considering $n=1$ (size of $u$'s similar users list), candidate sellers for $u$ in this graph would be $d$ and $e$.

## B. Computing Category Scores

Each item in the network belongs to a certain category. So we can calculate proximity of users according to the categories in which they have sold or bought items. I.e. we have a vector of non-numeric values for each user. There are bunch of algorithms for measuring similarity of such vectors. Here, the significant note is the importance of different categories for each user. For considering the importance of each category for each user, we used two factors:

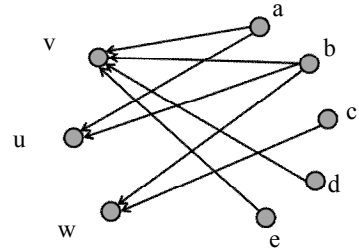

Fig. 2. *A sample commercial graph. Input edges imply buying and output edges imply selling items by users.*

1. The importance of each category in local network of a user which can be calculated using Eq. (2);

2. Number and value of items bought or sold in each category.

$$L_u(a) = \sum_{v \in N(u)} A(v, a) \qquad (2)$$

In Eq. (2), $L_u(a)$ is the importance of category $a$ in user's local network, $a$ is a sample category, and $N(u)$ is the list of $u$'s adjacent nodes. $A(v,a)$ is equal to 1 if both $u$ and $v$ have bought or sold an item in category of $a$, and 0 otherwise; If $u$ and $v$ does not have any category in common, $L_u(a)$ is equal to 1. Finally, we calculate ultimate score of each category for each user using Eq. (3) in which $Q_u(k)$ and $P_u(k)$ are respectively number and price of item $k$, and $L_u(a)$ is computed using Eq. (3).

$$w_u(a) = L_u(a) \times \sum_{k \in a} (Q_u(k) \times P_u(k)) \qquad (3)$$

Now we can use similarity measures like Jaccard's coefficient to compute the score of candidate sellers for the target user according to category of items. Eq. (4) is the adapted form of Jaccard's coefficient for our problem, Where $S_{category}(u,v)$ is the category score between $u$ and $v$, $c$ is a category that both $u$ and $v$ have traded in, $A(u)$ is the list of categories that user $u$ have traded in, and $w_u(a)$ is the category score of attribute $a$ according to user $u$:

$$S_{category}(u,v) = \sum_{c \in A(u) \cap A(v)} \frac{w_u(c) + w_v(c)}{\sum_{a \in A(u)} w_u(a) + \sum_{b \in A(v)} w_v(b)} \qquad (4)$$

### C. Computing Reputation Scores

Reputation is a quantity derived from the prior interactions of individuals to others in a community which is globally visible to all members of the network. Although the reputation of sellers is not the most important factor of increasing number of sales, but it is one of the effective factors for buying items from them [16], [17]. So we should consider the reputation of sellers in recommending items to target users. In commercial networks, the information about previous transactions can be used to calculate the reputation of users [18]. Generally, reputation systems are divided into two categories: centralized and distributed; In centralized reputation systems, a central authority is responsible for saving ratings or complaints and computing reputation of users, unlike distributed ones which there is no centralized authority for computing reputations [19]. The type of reputation system in our model is centralized but unlike common centralized reputation systems, every seller may have a different value of reputation form the viewpoint of different users. To compute reputation of each seller, four factors are considered: seller ratings from prior transactions, monetary value of each transaction, total number of seller's transactions, and importance of categories to target user. Eq.

(5) shows how we calculate a seller's reputation from perspective of a target user.

$$S_{reputation}(u,v) = \frac{\sum_{i=1}^{S(v)} (R_s(v,i) \times D(v,i) \times L(u, g(i)))}{Q(v)} \qquad (5)$$

$$\alpha + \beta = \chi. \qquad (1) \qquad (1)$$

$S_{reputation}(u,v)$ is the reputation of user $v$ from user $u$'s view. $Q(v)$ is total sales number of user $v$, $R_s(v,i)$ is the average value of item quality, responsiveness, shipment cost, and shipment time rating that seller $v$ have received in transaction $i$ (a number between [-1,1]). $D(v,i)$ is the monetary value of user $v$'s sales in transaction $i$. $L(u,g(i))$ is the importance of category in which user $v$ have sold an item and is computed similar to $L_u(a)$ in Eq. (2); If $u$ has not traded in category $g(i)$, $L(u,g(i))$ value will be equal to 1.

### D. Computing Ratings Scores

The other metric that can be used to find best sellers, is similarity between candidate sellers and prior sellers that have sold items to users, in terms of ratings they have gained. We assume that buyers rate sellers after each transaction, regarding to the overall satisfaction, quality of goods, delivery time, and seller support. These rating vectors can be used to choose best sellers for target users. Several metrics have been proposed in order to compute similarity of two numeric vectors. Cosine similarity [20] and Pearson correlation [21] are the most common ones. Here is the adopted version of cosine similarity we used to compute rating scores:

$$S_{rating}(u,v) = \sum_{w \in I(u)} \frac{\sum_{j \in R} (MR(v,j) \times MR(w,j))}{\sqrt{\sum_{j \in R} (MR(v,j))^2} \times \sqrt{\sum_{j \in R} (MR(w,j))^2}} \qquad (6)$$

$$\alpha + \beta = \chi.$$

In Eq. (6), $S_{rating}$ is the similarity of seller $v$ to sellers of user $u$ according to their ratings, $I(v)$ is the list of sellers that user $u$ has bought items from, $w$ is each user who has sold items to user $u$, $j$ is a rating component, and $R$ is total number of rating components (which is equal to 4 in our problem), $MR(v,j)$ is the average rating of user $v$ according to rating component $j$ in all of his sales, and $MR(w,j)$ is the average rating of user $w$ according to rating component $j$ in all of his sales.

### E. Computing Total Score of Candidates

The List of best sellers for a target user is built using three metrics we discussed before (Category score, Ratings score, and Reputation score). Before using these 3 scores, they should be normalized into a number between [0,1]. Eventually, total score of each candidate is calculated by combining the scores from previous stages:

$$S_{total}(u,v) = \alpha \times S_{cat}(u,v) + \beta \times S_{rat}(u,v) + \gamma \times S_{rep}(u,v) \qquad (7)$$

In Eq. (7), $S_{total}(u,v)$ is the total score of seller $v$ for recommending his items to user $u$, $\alpha$ is category score coefficient, $\beta$ is ratings score coefficient, $\gamma$ is reputation score coefficient, $S_{cat}(u,v)$ is category score of user $v$ according to user $u$, $S_{rat}(u,v)$ is rating score of user $v$ according to user $u$, and $S_{rep}(u,v)$ is reputation score of user $v$ according to user $u$. $\alpha$, $\beta$ and $\Upsilon$ must be chosen in range of $[0,1]$ so that $\alpha+\beta+\Upsilon=1$. By changing value of these coefficients, one can change the influence of each score in building best sellers list.

### F. Choosing items from sellers

After creating list of sellers for each user, we must choose one item from each seller's selling items. Online stores use several strategies for recommending items to users. These strategies include suggestion of random items, suggestion of best-selling items, using association rules (using algorithms like A-priori), collaborative filtering (using ratings of items), content based filtering (using textual information of items), demographic filtering (using age, gender, job, living area) and so on.

We could not use the last three methods listed above Hence users' demographic information, textual information about data, and ratings of users to items were not available in our dataset. So in our experiments, we used these methods to perform the last stage of our model: choosing best-selling items, choosing random items, and using association rules for item selection. In random item suggestion, a random item from selling items of a seller is selected and in Best-selling item suggestion, we select the best selling item of the seller. For applying association rules, we use the list of items that users have purchased so far. So we use a-priori algorithm to find strong rules and considering these rules, we select items from selling items each seller; If no rules could be used for item selection, we use random selection for suggesting items. The simplest form of recommendation is suggesting a single item. A single item would increase the chance that the customer will accept the suggestion. But most of the recommender systems provide a list of suggestions for customers. A list of items can increase the probability that user's desired item shows up in the list of items [22]. In our experiments, we select one item from each seller in order to shorten the final recommendation list.

### IV. DATA AND EXPERIMENTS

The best datasets that could be used for our experiments is the transaction data of users in C2C e-commerce websites. So we used a dataset of Bizrate.com to evaluate the performance of our proposed model. The dataset contains information about 2066 transactions between sellers and buyers. Each record of transaction includes: sellerID, buyerID, itemID, category, price, and ratings that seller has received from buyer. The ratings include four parameters: overall satisfaction, quality of goods, delivery time, and seller support. The graph of sellers-buyers graph has an average degree of 1.3, density of 0.001, maximum in-degree of 17 (i.e. purchases), and maximum out-degree of 167 (i.e. sales).

Precision and recall are the most common measures for evaluating prediction algorithms. Precision of a prediction algorithm is the number of truly predicted links divided by total number of predictions, and recall is the number of truly predicted links divided by total number of true links that can be predicted [12]. According to some studies in the field of information retrieval, as the level of precision increases, the level of recall decreases (and wise-versa). F-measure which is the harmonic mean of precision and recall, can be used to balance the two measures [23]. In order to compute precision, recall and F-measure, we used 10-fold cross validation [24]. So the data is partitioned into 10 nearly equally sized folds and 10 iterations of training and validation is performed. In each iteration, a different fold is held-out for validation (validation set) and the other 9 folds are used for learning (training set). The similarity of nodes and different scores (discussed in section III) are computed using the training set. The average value of precision, recall and F-measure obtained from these iterations is considered as estimated values of these measures on unseen data. To compute precision, recall and F-measure in each iteration 50 nodes that have least one edge in validation set are randomly selected and for each node, 1 to 25 predictions is made.

### V. RESULTS AND DISCUSSION

We measured precision, recall and F-measure at two levels: the first level, is to predict whether two users will show up in a same transaction in future or not (as buyer and seller); the second level is to predict sold items in future transactions. In other words, the first level is evaluating the model without performing stage 6. Obviously, size of precision, recall and F-measure of the model at stage 6 cannot be more than size of these metrics at stage 5. Fig. 3, Fig. 4 and Fig. 5 shows the percentage of precision, recall and F-measure of the model at user level (M1). In M2, the scores computed at stage 2, 3 and 4 are not used; instead, random sellers are selected from candidate list created at stage 1 to form prediction list for target users. As you see in Fig. 5, for M1, the level of F-measure decreases as the number of recommendations for each user increases. This can be a result of low number of missing link for each target user; in our experiments, this value was at least of 1 and at most 3 for each sample user. let Consider the number of missing edges of a sample node is equal to 2; If we assume that all of its missing links are truly predicted at the point of 5 prediction for the user, any other predictions for this user will decrease the overall precision of the system, because there is no more missing links for the user to be recovered.

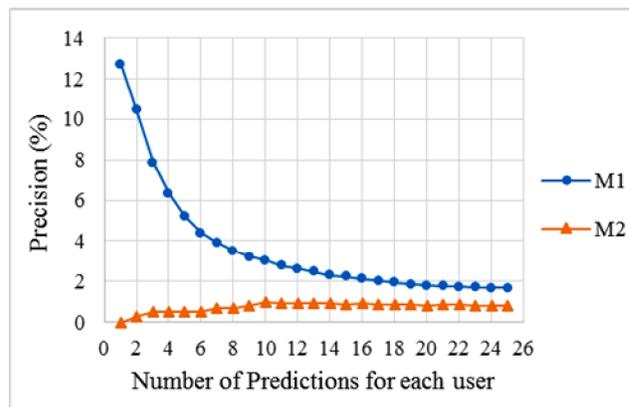

Fig. 3. *Precision (%) of proposed model. M1 and M2 respectively show the precision of model with and without using scores obtained from stage 2, 3 and 4 in fig. 1.*

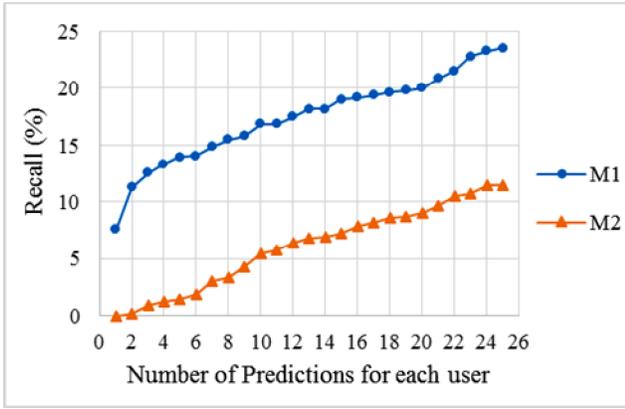

Fig. 4. *Recall (%) of proposed model. M1 and M2 respectively show the recall of model with and without using scores obtained from stage 2, 3 and 4 in fig. 1.*

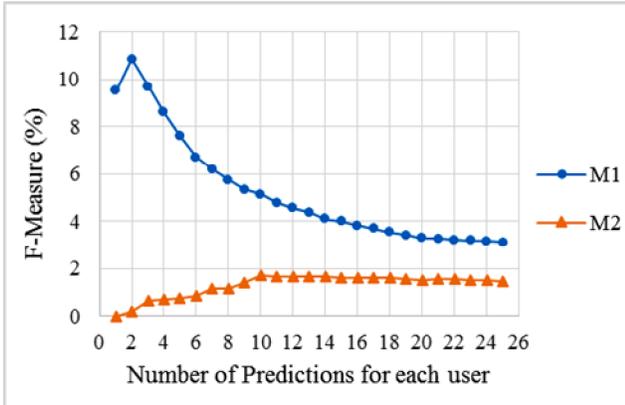

Fig. 5. *F-Measure of proposed model. M1 and M2 respectively show the F-Measure of model with and without using scores obtained from stage 2, 3 and 4 in fig. 1.*

The final form of the recommendation, is a list of items from different sellers, so we should measure the performance of model at predicting items too. Table I shows the f-measure of proposed model at item prediction using the three methods discussed in section III. As you see in the table, best-selling items method is having the best performance among others. But this method lacks personalization in recommending items to users. The performance of association rules method is being weak and almost close to random prediction method. The reason for this weak performance can be the weakness of extracted rules.

TABLE I.    MAXIMUM VALUE OF PRECISION, RECALL AND F-MEASURE FOR DIFFERENT METHODS OF ITEM SELECTION (STAGE 6 OF PROPOSED MODEL)

| Item Selection Method | Evaluation Metric (Maximum Value) | | |
|---|---|---|---|
| | *Precision* | *Recall* | *F-Measure* |
| Best-Selling Item | 10.79 | 25.34 | 10.44 |
| Association Rules | 0.15 | 0.49 | 0.21 |
| Random | 0.29 | 0.458 | 0.33 |

## VI. CONCLUSION AND FUTURE WORK

Online stores need an efficient e-marketing strategy. Modeling users' behavior and recommending items based on their interests is one of the most efficient strategies in e-commerce. Over time, recommender systems are becoming an inseparable component of e-commerce websites. But unfortunately, most of the current recommendation models could only be used in B2C e-commerce websites. In this paper, we proposed a recommender system for C2C e-commerce websites using prior transactions of users, categories of items, rating of users, and reputation of sellers. The goal of the system is recommending items from most appropriate sellers to target customers.

The Results of our experiments show that the proposed model can predict some of the missing links in commercial networks. The portion of predicted links by our model may seem to be low, but one should consider that even this low percent can end up to a big number of new transactions in large scale commercial networks. It should be noted that using different datasets for experiments may have significant effects on results, as the scale of our dataset was too large. Similar to other recommendation models, a challenge for our model is recommending items to new users (cold start). There are several ways to deal with this problem: recommending items from sellers who have sold most items in the network, sellers who have best reputation in the network and recommending items from random sellers are some common solutions for the problem.

The proposed model can be improved in several ways. The results of related works shows that considering time element in recommendations, can improve performance of recommender systems [10], [9]. So investigating the effect of using time in predictions is one of the future works. As the effect of using weight of links in a graph for link prediction is not deterministic yet [25], other possible future work can be using the weight of links between users for computing similarities at the stage of creating candidates list.

### REFERENCES

[1]  H. Wen, "Development of personalized online systems for web search, recommendations, and e-commerce," 2011.

[2]  A. Boroujeni, G. Assadat, and S. A. Hashemi Golpayegani, "Design Context Aware Recommender System Model in M-commerce Platform Using Collaborative Filtering Approach in Social Networks," *Caspian Journal of Applied Sciences Research,* vol. 2, 2013.

[3]  J. B. Schafer, J. A. Konstan, and J. Riedl, "E-commerce recommendation applications," in *Applications of Data Mining to Electronic Commerce,* ed: Springer, 2001, pp. 115-153.

[4]  F. Bonchi, C. Castillo, A. Gionis, and A. Jaimes, "Social network analysis and mining for business applications," *ACM Transactions on Intelligent Systems and Technology (TIST),* vol. 2, p. 22, 2011.

[5] D. Liben-Nowell and J. Kleinberg, "The link-prediction problem for social networks," *Journal of the American society for information science and technology,* vol. 58, pp. 1019-1031, 2007.

[6] H. R. de Sá and R. B. C. Prudêncio, "Supervised link prediction in weighted networks," in *Neural Networks (IJCNN), The 2011 International Joint Conference on,* 2011, pp. 2281-2288.

[7] C. Guangyao, "Research on the Recommending Method Used in C2C Online Trading," in *Web Intelligence and Intelligent Agent Technology Workshops, 2007 IEEE/WIC/ACM International Conferences on,* 2007, pp. 103-106.

[8] D. Ai, H. Zuo, and J. Yang, "C2C E-commerce Recommender System Based on Three-dimensional Collaborative Filtering," *Applied Mechanics and Materials,* vol. 336, pp. 2563-2566, 2013.

[9] O. Arazy, N. Kumar, and B. Shapira, "Improving social recommender systems," *IT professional,* vol. 11, pp. 38-44, 2009.

[10] I. Guy, N. Zwerdling, D. Carmel, I. Ronen, E. Uziel, S. Yogev, *et al.*, "Personalized recommendation of social software items based on social relations," in *Proceedings of the third ACM conference on Recommender systems,* 2009, pp. 53-60.

[11] L. Esmaeili, M. Nasiri, and B. Minaei-Bidgoli, "Personalizing group recommendation to social network users," in *Web Information Systems and Mining,* ed: Springer, 2011, pp. 124-133.

[12] Z. Yin, M. Gupta, T. Weninger, and J. Han, "A unified framework for link recommendation using random walks," in *Advances in Social Networks Analysis and Mining (ASONAM), 2010 International Conference on,* 2010, pp. 152-159.

[13] J. Chen, W. Geyer, C. Dugan, M. Muller, and I. Guy, "Make new friends, but keep the old: recommending people on social networking sites," in *Proceedings of the SIGCHI Conference on Human Factors in Computing Systems,* 2009, pp. 201-210.

[14] G. Jeh and J. Widom, "SimRank: a measure of structural-context similarity," in *Proceedings of the eighth ACM SIGKDD international conference on Knowledge discovery and data mining,* 2002, pp. 538-543.

[15] A. N. Langville and C. D. Meyer, *Google's PageRank and beyond: The science of search engine rankings*: Princeton University Press, 2011.

[16] P. Resnick and R. Zeckhauser, "Trust among strangers in Internet transactions: Empirical analysis of eBay's reputation system," *Advances in applied microeconomics,* vol. 11, pp. 127-157, 2002.

[17] L. Mui, M. Mohtashemi, and A. Halberstadt, "A computational model of trust and reputation," in *System Sciences, 2002. HICSS. Proceedings of the 35th Annual Hawaii International Conference on,* 2002, pp. 2431-2439.

[18] D. W. Manchala, "E-commerce trust metrics and models," *Internet Computing, IEEE,* vol. 4, pp. 36-44, 2000.

[19] A. Jøsang, R. Ismail, and C. Boyd, "A survey of trust and reputation systems for online service provision," *Decision support systems,* vol. 43, pp. 618-644, 2007.

[20] A. Rajaraman and J. D. Ullman, *Mining of massive datasets*: Cambridge University Press, 2012.

[21] B. Sarwar, G. Karypis, J. Konstan, and J. Riedl, "Item-based collaborative filtering recommendation algorithms," in *Proceedings of the 10th international conference on World Wide Web,* 2001, pp. 285-295.

[22] C.-T. Wu and H.-F. Wang, "Recent development of recommender systems," in *Industrial Engineering and Engineering Management, 2007 IEEE International Conference on,* 2007, pp. 228-232.

[23] M. K. Buckland and F. C. Gey, "The relationship between recall and precision," *JASIS,* vol. 45, pp. 12-19, 1994.

[24] P. Refaeilzadeh, L. Tang, and H. Liu, "Cross-validation," in *Encyclopedia of Database Systems,* ed: Springer, 2009, pp. 532-538.

[25] L. Lü and T. Zhou, "Link prediction in weighted networks: The role of weak ties," *EPL (Europhysics Letters),* vol. 89, p. 18001, 2010.